\documentclass[journal=nalefd]{achemso}
\usepackage{amsmath}

\title{Random Telegraph Signal in a Metallic Double-Dot System}

\author{Yuval Vardi}
\email{yuval.vardi@weizmann.ac.il}

\author{Avraham Guttman}
\author{Israel Bar-Joseph}

\affiliation{Department of Condensed Matter Physics, Weizmann Institute of Science, Rehovot 76100, Israel}

\begin{document}

\renewcommand*{\natmovechars}{}

\begin{abstract}
In this work we investigate the dynamics of a single electron surface trap, embedded in a self-assembly metallic double-dot system. The charging and discharging of the trap by a single electron is manifested as a random telegraph signal of the current through the double-dot device. We find that we can control the duration time that an electron resides in the trap through the current that flows in the device, between fractions of a second to more than an hour. We suggest that the observed switching is the electrical manifestation of the optical blinking phenomenon, commonly observed in semiconductor quantum dots.

Keywords: Random telegraph signal; Double dot; Metallic nanoparticles; Self-assembly; Charge trap; Photoluminescence blinking; Electrical transport;
\end{abstract}

Double quantum dot systems offer a unique opportunity for studying the world of quantum transport~\cite{VanderWiel2002}, and are considered promising candidates for quantum information operations~\cite{Loss1998}.
This stems from the ability to localize an electron in a limited region in space on the dot, and monitor its presence and properties, through the rich Coulomb blockade behavior~\cite{Kastner1992,Petta2005,Maune2012}.
This concept has been realized in a variety of material systems with controlled size and shape, down to a few nanometers scale~\cite{Banin1999a,feldheim2002metal,Bolotin2004, Hu2007, Todd2009, Guttman2011}.
Another system, in which electrons can be stored and measured, is an electronic trap in solid~\cite{Kane1998,Tyryshkin2003,Jelezko2004}.
The electrons in such a trap are better isolated from the environment. However, their measurement and control are more difficult.

Here we demonstrate how these two systems, metallic double-dots and electronic traps, are combined to yield a hybrid structure in which an electron can be stored for long durations and can be easily detected and measured. The charging and discharging of the trap by a single electron is manifested as a Random Telegraph Signal (RTS) of the current through the double-dot device.
RTS was studied in other systems of nano-objects, including semiconductor nanorods~\cite{Steinberg2010} and dots~\cite{Lachance-Quirion2014}, and molecular junctions~\cite{Neel2011}. It is commonly attributed to charging and discharging of surface traps, or to reconfiguration of molecular charge states.
Here we study RTS in a self-assembled metallic nanoparticle double-dot system. We show that we can control the duration that an electron resides in the trap through the current, varying it between fractions of a second to minutes. Furthermore, we show that at the Coulomb blockade region, when no current is flowing, the electron can be kept in the trap for extremely long times ($>$ hour).
We find significant similarities between the electrical RTS in the metallic nanoparticle system and the optical blinking observed in semiconductor quantum dots~\cite{Nirmal1996,Shimizu2001}, suggesting that these are the optical and electrical manifestation of the same phenomenon.

\begin{figure}
\centering\includegraphics[width=1\columnwidth]{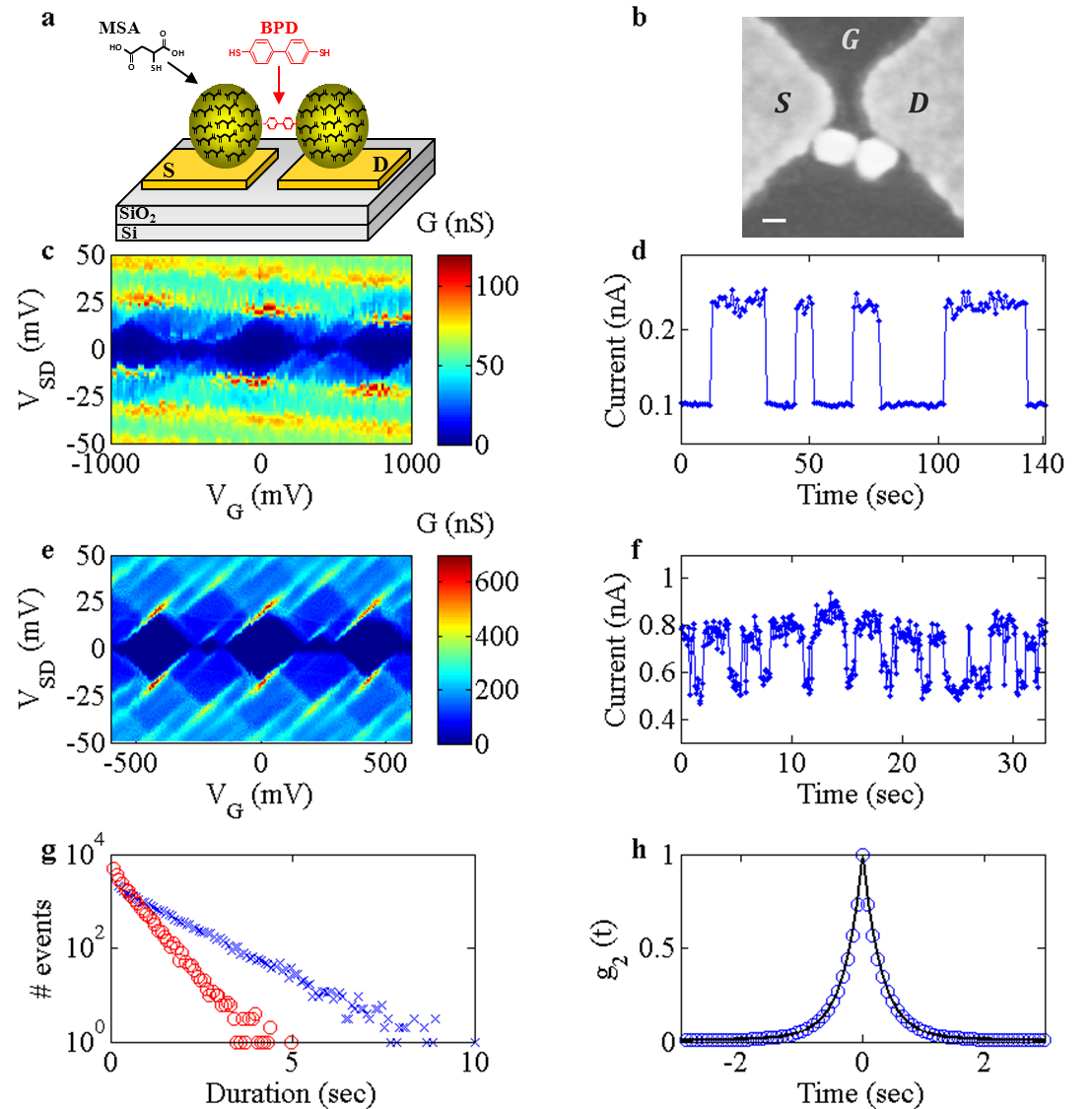}
\caption{
(a) Schematics of the metallic double-dot system. The dimer, composed of $34$ nm nano-spheres, is aligned in series with source $(S)$ and drain $(D)$ electrodes on top of a $100$ nm insulating SiO$_2$ layer. An n-doped Si substrate serves as a back-gate.
(b) Scanning electron microscope image of a trapped dimer. The scale bar corresponds to $20$ nm.
(c) Measurement of the differential conductance $(G=dI/dV_{SD})$ of device A as a function of the back-gate voltage $V_G$ and the source-drain bias voltage $V_{SD}$. A switching behavior is observed as jumps of the conductance peaks throughout the $V_G-V_{SD}$ plane.
(d) Random telegraph signal (RTS) of the current in device A ($V_{SD}=18$ mV, $V_G=0$ mV).
(e)-(f) The differential conductance and RTS ($V_{SD}=-16$ mV, $V_G=-400$ mV) measured in device B. The RTS is hardly seen in (e) due to the fast switching.
(g) A histogram of the lower (circles) and upper (X) states durations of the RTS presented in f. The exponential distribution of both states is evident, with $\tau_0=0.4$ sec and $\tau_1=1$ sec.
(h) Time auto-correlation function, $g_2\left(t\right)$, of the measured RTS presented in d (circles) and an exponential fit (solid line).
\label{fig1}}
\end{figure}

The system consists of two metallic nanoparticles (NPs) on top of a Si/SiO$_2$ substrate [Fig.~\ref{fig1}(a)-(b)]. The NPs are Au $34$ nm spheres, covered by a dense capping layer of mercaptosuccinic acid that acts as a tunnel barrier for electrical current into and from the NP. The dimer is formed by covalently linking the spheres with a short organic molecule, 4,4'-biphenyldithiol (BPD), thus forming a sub-nanometer gap between them~\cite{Dadosh2005,Guttman2011}, and placed between e-beam defined electrodes by means of electrostatic trapping~\bibnotemark[2]. The Si substrate is n-doped and used as a back gate. The measurements were done in a dilution refrigerator at a temperature of $60$ mK.

Typical electrical conductance properties as a function of the back gate voltage, $V_G$, and the source-drain voltage, $V_{SD}$, are shown in Fig.~\ref{fig1}(c),(e). The Coulomb blockade regime, consisting of a series of low conductance regions (diamonds) near zero $V_{SD}$ is clearly visible.
The large diamonds correspond to charging an extra electron on one of the dots, and the small ones - to charging the second dot while the first dot is already charged.

A noisy behavior is readily observed throughout the $V_G-V_{SD}$ plane of device A [Fig.~\ref{fig1}(c)]. We find that it corresponds to a time dependent switching behavior: Figure~\ref{fig1}(d) shows a measurement of the current $I$ for over $140$ sec. The current abruptly switches between $0.1$ nA to $0.25$ nA, and remains at each conductance state ("on" and "off") for time durations $t_0$ and $t_1$, which may vary in length. Such switching behavior, commonly known as random telegraph signal, is found in almost all devices we studied, each with different contrast and time durations. This "noisy" pattern is the focus of our study.

Figures~\ref{fig1}(e)-(f) show similar measurements of another device (device B), in which the conductance contrast is smaller and the time durations are substantially shorter. The fast dynamics of this device allows us to follow and analyze its behavior over many switching events. Figure~\ref{fig1}(g) shows a histogram of $t_{0,1}$ that represents more than $10^4$ switching events. Clear exponential distributions are seen, from which the characteristic lifetime of these states, $\tau_{0,1}$, can be extracted. Such exponential distributions are found in all measured devices.

The significance of these well-defined lifetimes can be realized by examining the time auto-correlation function of the measured RTS,
\begin{equation}
g_2\left(\Delta t\right)=\frac{\left\langle \tilde{I}\left(t\right) \tilde{I}\left(t+\Delta t\right) \right\rangle}{\left\langle \tilde{I}^2\left(t\right) \right\rangle},\label{eq:g2_def}
\end{equation}
where $\tilde{I}\left(t\right)=I\left(t\right)-\left\langle I \right\rangle$, and $\left\langle \cdot \right\rangle$ represents time average. We find that this function falls exponentially at a decay rate $\Gamma$ which is the sum of the two inverse lifetimes, $\Gamma=\tau_0^{-1}+\tau_1^{-1}$ [Fig.~\ref{fig1}(h)]. This behavior provides a useful insight into the underlying mechanism of RTS, as it describes a trap that is fed by an electron at a rate $\gamma_0=\tau_0^{-1}$ and is emptied at $\gamma_1=\tau_1^{-1}$.
One can describe the rate equations for its occupation probabilities $P_{0,1}$ as:
\begin{subequations}
	\label{eq:rate_equations}
	\begin{eqnarray}
		\frac{dP_1}{dt}&=&\gamma_0P_0\left(t\right)-\gamma_1P_1\left(t\right),
		\\
		\frac{dP_0}{dt}&=&-\gamma_0P_0\left(t\right)+\gamma_1P_1\left(t\right),
	\end{eqnarray}
\end{subequations}
where $P_0\left(t\right)+P_1\left(t\right)=1$. Solving equations~\eqref{eq:rate_equations} we indeed get
$g_2\left(\Delta t\right)=\exp\left(-\Gamma\cdot\Delta t\right)$. We, therefore, conclude that a trapped electron gates the double dot system and modifies its electrical conductance: different current levels correspond to different states of the trap - either empty or occupied.

\begin{figure}
\centering\includegraphics[width=1\columnwidth]{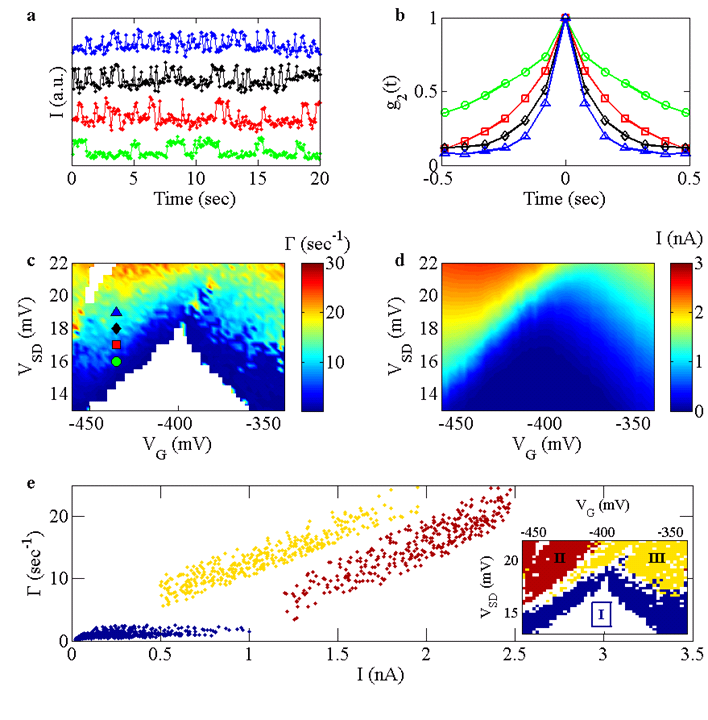}
\caption{
(a) RTS and (b) the corresponding time auto-correlation $g_2\left(t\right)$ for $V_{SD}$ values between $16-19$ mV. The increase of the decay rate $\Gamma$ with voltage is clearly seen.
(c) The behavior of $\Gamma$ across the $V_G-V_{SD}$ plane. The blank area at the bottom is the Coulomb blockade region, where RTS cannot be detected. The markers indicate the values at which the measurements shown in a,b were conducted.
(d) The behavior of the current across the $V_G-V_{SD}$ plane, demonstrating a similar dependence as $\Gamma$ in c.
(e) $\Gamma$ as a function of the current, revealing the existence of 3 distinct clusters (I-blue, II-red, III-yellow). The inset shows the location of these clusters in the $V_G-V_{SD}$ plane (same colors).
\label{fig2}}
\end{figure}

We map the dependence of the decay rate, $\Gamma$, across the $V_G-V_{SD}$ plane: for each $\left(V_G,V_{SD}\right)$ value RTS was measured [Fig.~\ref{fig2}(a)], the corresponding $g_2$ was calculated [Fig.~\ref{fig2}(b)], and $\Gamma$ was extracted [Fig.~\ref{fig2}(c)]. We find that $\Gamma$ is slower near the Coulomb blockade region, and becomes faster away from it. This behavior of $\Gamma$ nicely mirrors the current, $I$, in the $V_G-V_{SD}$ plane [Fig.~\ref{fig2}(d)], indicating that the charging and discharging process of the trap is governed by the attempt rate of the flowing electrons.
Figure~\ref{fig2}(e) shows $\Gamma$ as a function of $I$ for $1500$ points in the $V_G-V_{SD}$ plane. Indeed, $\Gamma$ increases linearly with the current throughout, except for the Coulomb blockage region which will be discusses later.

We find a clear clustering of the measured decay rates into three groups [Fig.~\ref{fig2}(e)].
The physical origin of this clustering can be understood by calculating the average charging state of the two dots at each $\left(V_G,V_{SD}\right)$ value~\bibnotemark[2]. We identify three main regions: the blockade [marked as I in Fig.~\ref{fig2}(e)], in which the charging state is $\left(n_1=-1, n_2=-1\right)$, the upper left region (II), in which an additional state, $(-2,-1)$, contributes to the conduction, and the upper right region (III), with $(-1,0)$ state. Here, $n_i$ is the number of excess electrons on the i'th dot.
The clustering of $\Gamma$ follows these charging states: the two fast switching clusters (II and III) have the same average charge difference, $\delta n=n_1-n_2$, and differ in the average total charge, $\Sigma n=n_1+n_2$, while the slow region near the blockade (region I) is characterized by $\delta n=0$. We conclude that the trap dynamics is determined by the charging state of the two dots such that the electrostatic field between the dots, $\delta n$, acts as an effective drain-source field and the electrostatic potential, $\Sigma n$, acts as an effective gate.

\begin{figure}
\centering\includegraphics[width=1\columnwidth]{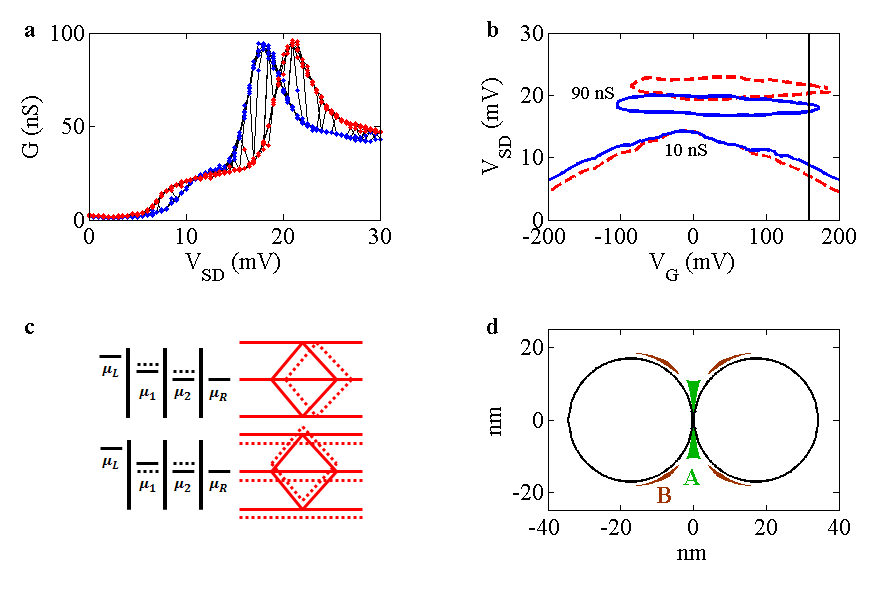}
\caption{
(a) Multiple conductance scans of device A as a function of $V_{SD}$ at $V_G=160$ mV. The red and blue points depict the "on" and "off" states, respectively. In each state the spectrum consists of two peaks which are shifted relative to each other in opposite directions.
(b) Contour plot of the conductance peaks of the two states across the $V_G-V_{SD}$ plane. The contour lines shift to opposite directions along the $V_{SD}$ axis (vertically). The solid vertical black line represents the measurement shown in a.
(c) Diagrams of the electrochemical potentials of the leads $\left(\mu_L, \mu_R\right)$ and the dots $\left(\mu_1, \mu_2\right)$ for the case of common gating ($\delta\mu_1=\delta\mu_2$, top left) and differential gating ($\delta\mu_1=-\delta\mu_2$, bottom left). The schemes at the right show the conductance peaks in the $V_G-V_{SD}$ plane. It is seen that common gating causes a horizontal shift in the conductance spectrum (top right), while differential gating causes a vertical shift of the peaks in opposite directions (bottom right).
(d) The area where the trap can be located in device A (green, mainly differential gating: $\delta\mu_d>0.1E_C$, $\delta\mu_c<0.02E_C$) and B (brown, mixing of common and differential gating: $0.02E_C<\delta\mu_d,\delta\mu_c<0.05E_C$), as calculated in the electrostatic numerical simulation. Here $\delta\mu_d,\delta\mu_c$ are the differential and common gating, respectively.
\label{fig3}}
\end{figure}

By examining the conductance, $G$, we prove that the trap is indeed located between the dots near their surface. Figure~\ref{fig3}(a) shows multiple $G-V_{SD}$ scans of device A for a fixed $V_G$ value. This device exhibits a large RTS contrast [Fig.~\ref{fig1}(d)], and, hence, it is easier to follow its conductance properties. It is seen that each conducting state ("on" or "off") is characterized by a distinct conductance spectrum, and the system jumps randomly from one to another. Hence, by measuring the conductance spectra for various $V_G$ values we can map the location of the conductance peaks of the two states in the $V_G-V_{SD}$ plane [Fig.~\ref{fig3}(b)]. We find that device A exhibits a rather surprising behavior, where the conductance peaks are shifted vertically with respect to each other.

Clearly, a trapped electron in the vicinity of the dimer shifts the two electrochemical potentials, and should induce a horizontal shift of the conductance spectrum in the $V_G-V_{SD}$ plane [Fig.~\ref{fig3}(c), top panel]. A significant vertical shift can only be obtained if an electron moves from one of the dots into a trap that is located between the dots. This process lowers the electrochemical potential of the first dot by $\delta\mu_1=-\left(E_C-E_{Cm}\right)/2$, and raises the other by $\delta\mu_2=-\delta\mu_1$  [Fig.~\ref{fig3}(c), lower panel]. Here $E_C$ and $E_{Cm}$ are the individual and mutual charging energies of the dots, respectively.

We conduct a full numerical calculation of the induced gating of the two dots as a function of the trap location~\bibnotemark[2]. This allows us to locate the traps in devices A and B as shown in Fig.~\ref{fig3}(d). We find that the high contrast and vertical shift in device A are, indeed, due to a trap that is located in a rather small area, confined to the inter-dot region. The situation is somehow different for device B. Here we find a combination of horizontal and vertical shifts, with a smaller magnitude. This is consistent with a trap that is located slightly away from the central region between the dots, as depicted in Fig.~\ref{fig3}(d). Indeed, we find a range of relative shifts in the seven measured devices, indicating that the trap may appear in different locations near the inter-dot region, restricted to a layer close to the surface of the dots~\bibnote{A surface charge that is far away from the central region will not cause any significant effect, since it is equivalent to charging the dot by a full electron charge.}.

These location measurements suggest that the trap is related to the surrounding environment around the nanoparticles, i.e. the linking molecule or the capping layer. Indeed, the use of metallic nanoparticles allows us to rule out any trap within the nanoparticles volume.
The linking molecule can be easily eliminated as the source for the trap - we studied the conductance of spontaneously formed dimers which did not contain the linking BPD molecule~\bibnotemark[2], and observed a similar RTS behavior. The RTS signal was also found to be present in dimers which have a different capping layer (Na$_3$citrate)~\cite{Dadosh2005}, suggesting that the phenomenon is rather general and independent of the particular chemical composition of the layer.
One may consider two types of traps that may give rise to the observed behavior: a region within the capping layer in which an electron may be localized, or a reconfiguration of a polar molecule, which is effectively equivalent to a movement of a charge near the nanoparticles~\cite{Schirm2013}. Further studies are needed to clarify the exact nature of the trap.

\begin{figure}
\centering\includegraphics[width=1\columnwidth]{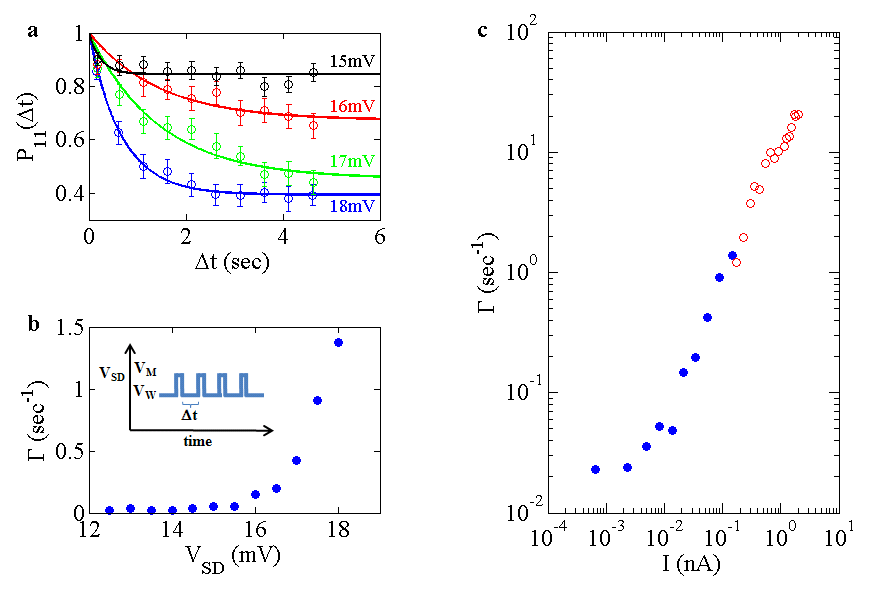}
\caption{
(a) The conditional probability $P_{11}\left(\Delta t\right)$ for the system to remain in the same state after wait time of $\Delta t$ in various $V_{SD}$ values.
(b) The corresponding decay rate $\Gamma$ as a function of $V_{SD}$. These measurements were performed as depicted at the inset of b, measuring the current at $V_{SD}=V_M$, and waiting at $V_{SD}=V_W$ for time duration $\Delta t$. We kept $V_M=18$ mV throughout, and varied $V_W$ between $12-18$ mV, well into the Coulomb blockade regime.
(c) $\Gamma$ as a function of the current in a log-log scale, inside (blue dots) and outside (red circles) the Coulomb blockade region. The blue dots are the ones presented in b, while the red ones - in Fig.~\ref{fig2}. Remarkably, the slope of the curve is approximately $1$, indicating a linear dependence of $\Gamma$ as a function of current for over $3$ orders of magnitude.
\label{fig4}}
\end{figure}

This system, a dimer with a charged trap, offers unique opportunities to control a single electron on macroscopic time scales and easily measure its state and dynamics. This control is demonstrated in Fig.~\ref{fig4}, which shows that an electron can be trapped for extremely long times in the Coulomb blockade region. To extract $\Gamma$ in this region, where the current is very small and RTS cannot be detected, we performed the following sequence: We measured the conductance at a $V_{SD}$ value, $V_M$, outside the blockade and determine whether it is in the "on" or "off" state. We then switched the voltage into a lower value in the blockade, $V_W$, and waited for a time duration $\Delta t$ at that voltage. Finally, we switched the voltage back to $V_M$ and measured again the conductance value [see inset of Fig.~\ref{fig4}(b)]. This allowed us to determine the conditional probability $P_{11}\left(\Delta t\right)$ to find the trap in the "on" state at time $t+\Delta t$ if it was in the same state at time $t$. Figure~\ref{fig4}(a) shows this conditional probability for various values of $V_W$. It is seen that as we lower $V_W$ and enter the blockade region the rate at which $P_{11}$ decays with $\Delta t$ becomes slower until it is nearly flat.
Extracting the decay rate, $\Gamma$, from these measurements, we find that it falls fast as we enter the Coulomb blockade regime and the occupation time of the trap exceeds a minute [Fig.~\ref{fig4}(b)].

Combining the decay rates measured in the blockade region with the previous ones, outside the blockade [presented in Fig.~\ref{fig2}(c)], we find a nice matching of the two sets of measurements [Fig.~\ref{fig4}(c)]. It is seen that $\Gamma$ depends linearly on $I$ over several orders of magnitude: the slope of $\Gamma$ versus $I$, plotted on a log-log scale, is approximately $1$. We repeated this measurement with $V_W=0$ and found no decay of the probability $P_{11}\left(\Delta t\right)$ up to $\Delta t=10$ minutes, implying that $\Gamma$ is much longer at this voltage~\bibnotemark[2]. Since no current is flowing this isolated electron has limited interaction with the environment.

A clear demonstration of a single electron operation that can be realized in this system is the ability to 'write' the trap state to the "on" or "off" position, and keep it for a desired duration~\bibnotemark[2]. This is done by varying the applied voltages between two values. The 'readout' is simply done by measuring the current through the device.

The characteristics of RTS in our system have many similarities with the optical blinking observed in semiconductor quantum dots~\cite{Nirmal1996,Shimizu2001} and wires~\cite{Glennon2007}, in particular the abrupt switching between two distinct states, the relatively long time scales (order of seconds), and the current (intensity) dependence~\cite{Banin1999,Pistol1999}. Furthermore, it is commonly accepted that optical blinking in these systems is due to a trapped charge at the surface~\cite{Efros1997}, and there is a strong experimental evidence for a gating effect caused by this trapped charge~\cite{Shimizu2002,Park2007}. These similarities suggest that the RTS observed in our work is the electrical manifestation of optical blinking, and can be used as a powerful tool to study this important effect.

We note, however, that there is a significant difference in the temporal statistics. While in our system the distribution of the "on" and "off" durations is exponential, the behavior in semiconductor dots is characterized by a power-law distribution~\cite{Shimizu2001}. In fact, the first quantum dot blinking model, which attributed the phenomenon to a trapping of carrier on the surface, predicted an exponential distribution of the switching times~\cite{Efros1997}. The origin of the power-law behavior remained not fully understood despite intense research in the last two decades~\cite{Frantsuzov2008}. A series of modifications to the surface trap model have been proposed, in particular - the existence of multiple electron traps, varying with distance and/or trap depth, near the quantum dot surface~\cite{Verberk2002}. If this is indeed the origin of the power-law behavior, one can understand why it is not relevant in our system, where the effective traps are limited to a narrow area in the inter-dot region. We suggest that measurements of the electrical conductance of semiconductor quantum dots may be instrumental in deciphering the lingering question of optical blinking.

\begin{suppinfo}
Nanoparticles synthesis and device fabrication methods, double-dot simulation, theoretical analysis of a trap dynamics, further measurements in the Coulomb blockade and demonstration of control, numerical calculations of the trap gating.
\end{suppinfo}

The authors declare no competing financial interest.

\begin{acknowledgement}
We would like to thank D. Mahalu and O. Raslin for their help in the electron beam lithography, and E. Cohen-Hoshen and J. Sperling for fruitful discussions on fabrication techniques.
\end{acknowledgement}

\bibnotetext[2]{See Supporting Information}

\bibliography{blinking}

\providecommand{\latin}[1]{#1}
\providecommand*\mcitethebibliography{\thebibliography}
\csname @ifundefined\endcsname{endmcitethebibliography}
  {\let\endmcitethebibliography\endthebibliography}{}
\begin{mcitethebibliography}{32}
\providecommand*\natexlab[1]{#1}
\providecommand*\mciteSetBstSublistMode[1]{}
\providecommand*\mciteSetBstMaxWidthForm[2]{}
\providecommand*\mciteBstWouldAddEndPuncttrue
  {\def\EndOfBibitem{\unskip.}}
\providecommand*\mciteBstWouldAddEndPunctfalse
  {\let\EndOfBibitem\relax}
\providecommand*\mciteSetBstMidEndSepPunct[3]{}
\providecommand*\mciteSetBstSublistLabelBeginEnd[3]{}
\providecommand*\EndOfBibitem{}
\mciteSetBstSublistMode{f}
\mciteSetBstMaxWidthForm{subitem}{(\alph{mcitesubitemcount})}
\mciteSetBstSublistLabelBeginEnd
  {\mcitemaxwidthsubitemform\space}
  {\relax}
  {\relax}

\bibitem[van~der Wiel \latin{et~al.}(2002)van~der Wiel, {De Franceschi},
  Elzerman, Fujisawa, Tarucha, and Kouwenhoven]{VanderWiel2002}
van~der Wiel,~W.; {De Franceschi},~S.; Elzerman,~J.~M.; Fujisawa,~T.;
  Tarucha,~S.; Kouwenhoven,~L.~P. \emph{Rev. Mod. Phys.} \textbf{2002},
  \emph{75}, 1--22\relax
\mciteBstWouldAddEndPuncttrue
\mciteSetBstMidEndSepPunct{\mcitedefaultmidpunct}
{\mcitedefaultendpunct}{\mcitedefaultseppunct}\relax
\EndOfBibitem
\bibitem[Loss and DiVincenzo(1998)Loss, and DiVincenzo]{Loss1998}
Loss,~D.; DiVincenzo,~D.~P. \emph{Phys. Rev. A} \textbf{1998}, \emph{57},
  120--126\relax
\mciteBstWouldAddEndPuncttrue
\mciteSetBstMidEndSepPunct{\mcitedefaultmidpunct}
{\mcitedefaultendpunct}{\mcitedefaultseppunct}\relax
\EndOfBibitem
\bibitem[Kastner(1992)]{Kastner1992}
Kastner,~M. \emph{Rev. Mod. Phys.} \textbf{1992}, \emph{64}, 849--858\relax
\mciteBstWouldAddEndPuncttrue
\mciteSetBstMidEndSepPunct{\mcitedefaultmidpunct}
{\mcitedefaultendpunct}{\mcitedefaultseppunct}\relax
\EndOfBibitem
\bibitem[Petta \latin{et~al.}(2005)Petta, Johnson, Taylor, Laird, Yacoby,
  Lukin, Marcus, Hanson, and Gossard]{Petta2005}
Petta,~J.~R.; Johnson,~A.~C.; Taylor,~J.~M.; Laird,~E.~A.; Yacoby,~A.;
  Lukin,~M.~D.; Marcus,~C.~M.; Hanson,~M.~P.; Gossard,~A.~C. \emph{Science (New
  York, N.Y.)} \textbf{2005}, \emph{309}, 2180--4\relax
\mciteBstWouldAddEndPuncttrue
\mciteSetBstMidEndSepPunct{\mcitedefaultmidpunct}
{\mcitedefaultendpunct}{\mcitedefaultseppunct}\relax
\EndOfBibitem
\bibitem[Maune \latin{et~al.}(2012)Maune, Borselli, Huang, Ladd, Deelman,
  Holabird, Kiselev, Alvarado-Rodriguez, Ross, Schmitz, Sokolich, Watson,
  Gyure, and Hunter]{Maune2012}
Maune,~B.~M.; Borselli,~M.~G.; Huang,~B.; Ladd,~T.~D.; Deelman,~P.~W.;
  Holabird,~K.~S.; Kiselev,~A.~A.; Alvarado-Rodriguez,~I.; Ross,~R.~S.;
  Schmitz,~A.~E.; Sokolich,~M.; Watson,~C.~A.; Gyure,~M.~F.; Hunter,~A.~T.
  \emph{Nature} \textbf{2012}, \emph{481}, 344--7\relax
\mciteBstWouldAddEndPuncttrue
\mciteSetBstMidEndSepPunct{\mcitedefaultmidpunct}
{\mcitedefaultendpunct}{\mcitedefaultseppunct}\relax
\EndOfBibitem
\bibitem[Banin \latin{et~al.}(1999)Banin, Cao, Katz, and Millo]{Banin1999a}
Banin,~U.; Cao,~Y.; Katz,~D.; Millo,~O. \emph{Nature} \textbf{1999},
  \emph{400}, 542--544\relax
\mciteBstWouldAddEndPuncttrue
\mciteSetBstMidEndSepPunct{\mcitedefaultmidpunct}
{\mcitedefaultendpunct}{\mcitedefaultseppunct}\relax
\EndOfBibitem
\bibitem[Feldheim and Foss(2002)Feldheim, and Foss]{feldheim2002metal}
Feldheim,~D.~L.; Foss,~C.~A. In \emph{{Metal Nanoparticles: Synthesis,
  Characterization, and Applications}}, illustrate ed.; Feldheim,~D.~L.,
  Foss,~C.~A., Eds.; Marcel Dekker, 2002; p 338\relax
\mciteBstWouldAddEndPuncttrue
\mciteSetBstMidEndSepPunct{\mcitedefaultmidpunct}
{\mcitedefaultendpunct}{\mcitedefaultseppunct}\relax
\EndOfBibitem
\bibitem[Bolotin \latin{et~al.}(2004)Bolotin, Kuemmeth, Pasupathy, and
  Ralph]{Bolotin2004}
Bolotin,~K.~I.; Kuemmeth,~F.; Pasupathy,~A.~N.; Ralph,~D.~C. \emph{Appl. Phys.
  Lett.} \textbf{2004}, \emph{84}, 3154--3156\relax
\mciteBstWouldAddEndPuncttrue
\mciteSetBstMidEndSepPunct{\mcitedefaultmidpunct}
{\mcitedefaultendpunct}{\mcitedefaultseppunct}\relax
\EndOfBibitem
\bibitem[Hu \latin{et~al.}(2007)Hu, Churchill, Reilly, Xiang, Lieber, and
  Marcus]{Hu2007}
Hu,~Y.; Churchill,~H. O.~H.; Reilly,~D.~J.; Xiang,~J.; Lieber,~C.~M.;
  Marcus,~C.~M. \emph{Nat. Nanotechnol.} \textbf{2007}, \emph{2}, 622--5\relax
\mciteBstWouldAddEndPuncttrue
\mciteSetBstMidEndSepPunct{\mcitedefaultmidpunct}
{\mcitedefaultendpunct}{\mcitedefaultseppunct}\relax
\EndOfBibitem
\bibitem[Todd \latin{et~al.}(2009)Todd, Chou, Amasha, and
  Goldhaber-Gordon]{Todd2009}
Todd,~K.; Chou,~H.-T.; Amasha,~S.; Goldhaber-Gordon,~D. \emph{Nano Lett.}
  \textbf{2009}, \emph{9}, 416--21\relax
\mciteBstWouldAddEndPuncttrue
\mciteSetBstMidEndSepPunct{\mcitedefaultmidpunct}
{\mcitedefaultendpunct}{\mcitedefaultseppunct}\relax
\EndOfBibitem
\bibitem[Guttman \latin{et~al.}(2011)Guttman, Mahalu, Sperling, Cohen-Hoshen,
  and Bar-Joseph]{Guttman2011}
Guttman,~A.; Mahalu,~D.; Sperling,~J.; Cohen-Hoshen,~E.; Bar-Joseph,~I.
  \emph{Appl. Phys. Lett.} \textbf{2011}, \emph{99}, 063113\relax
\mciteBstWouldAddEndPuncttrue
\mciteSetBstMidEndSepPunct{\mcitedefaultmidpunct}
{\mcitedefaultendpunct}{\mcitedefaultseppunct}\relax
\EndOfBibitem
\bibitem[Kane(1998)]{Kane1998}
Kane,~B.~E. \emph{Nature} \textbf{1998}, \emph{393}, 133--137\relax
\mciteBstWouldAddEndPuncttrue
\mciteSetBstMidEndSepPunct{\mcitedefaultmidpunct}
{\mcitedefaultendpunct}{\mcitedefaultseppunct}\relax
\EndOfBibitem
\bibitem[Tyryshkin \latin{et~al.}(2003)Tyryshkin, Lyon, Astashkin, and
  Raitsimring]{Tyryshkin2003}
Tyryshkin,~A.~M.; Lyon,~S.~A.; Astashkin,~A.~V.; Raitsimring,~A.~M. \emph{Phys.
  Rev. B} \textbf{2003}, \emph{68}, 193207\relax
\mciteBstWouldAddEndPuncttrue
\mciteSetBstMidEndSepPunct{\mcitedefaultmidpunct}
{\mcitedefaultendpunct}{\mcitedefaultseppunct}\relax
\EndOfBibitem
\bibitem[Jelezko \latin{et~al.}(2004)Jelezko, Gaebel, Popa, Gruber, and
  Wrachtrup]{Jelezko2004}
Jelezko,~F.; Gaebel,~T.; Popa,~I.; Gruber,~A.; Wrachtrup,~J. \emph{Phys. Rev.
  Lett.} \textbf{2004}, \emph{92}, 076401\relax
\mciteBstWouldAddEndPuncttrue
\mciteSetBstMidEndSepPunct{\mcitedefaultmidpunct}
{\mcitedefaultendpunct}{\mcitedefaultseppunct}\relax
\EndOfBibitem
\bibitem[Steinberg \latin{et~al.}(2010)Steinberg, Wolf, Faust, Salant, Lilach,
  Millo, and Banin]{Steinberg2010}
Steinberg,~H.; Wolf,~O.; Faust,~A.; Salant,~A.; Lilach,~Y.; Millo,~O.;
  Banin,~U. \emph{Nano Lett.} \textbf{2010}, \emph{10}, 2416--20\relax
\mciteBstWouldAddEndPuncttrue
\mciteSetBstMidEndSepPunct{\mcitedefaultmidpunct}
{\mcitedefaultendpunct}{\mcitedefaultseppunct}\relax
\EndOfBibitem
\bibitem[Lachance-Quirion \latin{et~al.}(2014)Lachance-Quirion, Tremblay,
  Lamarre, M\'{e}thot, Gingras, {Camirand Lemyre}, Pioro-Ladri\`{e}re, and
  Allen]{Lachance-Quirion2014}
Lachance-Quirion,~D.; Tremblay,~S.; Lamarre,~S.~A.; M\'{e}thot,~V.;
  Gingras,~D.; {Camirand Lemyre},~J.; Pioro-Ladri\`{e}re,~M.; Allen,~C.~N.
  \emph{Nano Lett.} \textbf{2014}, \emph{14}, 882--887\relax
\mciteBstWouldAddEndPuncttrue
\mciteSetBstMidEndSepPunct{\mcitedefaultmidpunct}
{\mcitedefaultendpunct}{\mcitedefaultseppunct}\relax
\EndOfBibitem
\bibitem[Neel \latin{et~al.}(2011)Neel, Kroger, and Berndt]{Neel2011}
Neel,~N.; Kroger,~J.; Berndt,~R. \emph{Nano Lett.} \textbf{2011}, \emph{11},
  3593--6\relax
\mciteBstWouldAddEndPuncttrue
\mciteSetBstMidEndSepPunct{\mcitedefaultmidpunct}
{\mcitedefaultendpunct}{\mcitedefaultseppunct}\relax
\EndOfBibitem
\bibitem[Nirmal \latin{et~al.}(1996)Nirmal, Dabbousi, Bawendi, Macklin,
  Trautman, Harris, and Brus]{Nirmal1996}
Nirmal,~M.; Dabbousi,~B.~O.; Bawendi,~M.~G.; Macklin,~J.~J.; Trautman,~J.~K.;
  Harris,~T.~D.; Brus,~L.~E. \emph{Nature} \textbf{1996}, \emph{383},
  802--804\relax
\mciteBstWouldAddEndPuncttrue
\mciteSetBstMidEndSepPunct{\mcitedefaultmidpunct}
{\mcitedefaultendpunct}{\mcitedefaultseppunct}\relax
\EndOfBibitem
\bibitem[Shimizu \latin{et~al.}(2001)Shimizu, Neuhauser, Leatherdale,
  Empedocles, Woo, and Bawendi]{Shimizu2001}
Shimizu,~K.~T.; Neuhauser,~R.~G.; Leatherdale,~C.~A.; Empedocles,~S.~A.;
  Woo,~W.~K.; Bawendi,~M.~G. \emph{Phys. Rev. B} \textbf{2001}, \emph{63},
  205316\relax
\mciteBstWouldAddEndPuncttrue
\mciteSetBstMidEndSepPunct{\mcitedefaultmidpunct}
{\mcitedefaultendpunct}{\mcitedefaultseppunct}\relax
\EndOfBibitem
\bibitem[Dadosh \latin{et~al.}(2005)Dadosh, Gordin, Krahne, Khivrich, Mahalu,
  Frydman, Sperling, Yacoby, and Bar-Joseph]{Dadosh2005}
Dadosh,~T.; Gordin,~Y.; Krahne,~R.; Khivrich,~I.; Mahalu,~D.; Frydman,~V.;
  Sperling,~J.; Yacoby,~A.; Bar-Joseph,~I. \emph{Nature} \textbf{2005},
  \emph{436}, 677--80\relax
\mciteBstWouldAddEndPuncttrue
\mciteSetBstMidEndSepPunct{\mcitedefaultmidpunct}
{\mcitedefaultendpunct}{\mcitedefaultseppunct}\relax
\EndOfBibitem
\bibitem[2()]{2}
See Supporting Information\relax
\mciteBstWouldAddEndPuncttrue
\mciteSetBstMidEndSepPunct{\mcitedefaultmidpunct}
{\mcitedefaultendpunct}{\mcitedefaultseppunct}\relax
\EndOfBibitem
\bibitem[Not()]{Note-4}
A surface charge that is far away from the central region will not cause any
  significant effect, since it is equivalent to charging the dot by a full
  electron charge.\relax
\mciteBstWouldAddEndPunctfalse
\mciteSetBstMidEndSepPunct{\mcitedefaultmidpunct}
{}{\mcitedefaultseppunct}\relax
\EndOfBibitem
\bibitem[Schirm \latin{et~al.}(2013)Schirm, Matt, Pauly, Cuevas, Nielaba, and
  Scheer]{Schirm2013}
Schirm,~C.; Matt,~M.; Pauly,~F.; Cuevas,~J.~C.; Nielaba,~P.; Scheer,~E.
  \emph{Nat. Nanotechnol.} \textbf{2013}, \emph{8}, 645--8\relax
\mciteBstWouldAddEndPuncttrue
\mciteSetBstMidEndSepPunct{\mcitedefaultmidpunct}
{\mcitedefaultendpunct}{\mcitedefaultseppunct}\relax
\EndOfBibitem
\bibitem[Glennon \latin{et~al.}(2007)Glennon, Tang, Buhro, and
  Loomis]{Glennon2007}
Glennon,~J.~J.; Tang,~R.; Buhro,~W.~E.; Loomis,~R.~a. \emph{Nano Lett.}
  \textbf{2007}, \emph{7}, 3290--5\relax
\mciteBstWouldAddEndPuncttrue
\mciteSetBstMidEndSepPunct{\mcitedefaultmidpunct}
{\mcitedefaultendpunct}{\mcitedefaultseppunct}\relax
\EndOfBibitem
\bibitem[Banin \latin{et~al.}(1999)Banin, Bruchez, Alivisatos, Ha, Weiss, and
  Chemla]{Banin1999}
Banin,~U.; Bruchez,~M.; Alivisatos,~a.~P.; Ha,~T.; Weiss,~S.; Chemla,~D.~S.
  \emph{J. Chem. Phys.} \textbf{1999}, \emph{110}, 1195--1201\relax
\mciteBstWouldAddEndPuncttrue
\mciteSetBstMidEndSepPunct{\mcitedefaultmidpunct}
{\mcitedefaultendpunct}{\mcitedefaultseppunct}\relax
\EndOfBibitem
\bibitem[Pistol \latin{et~al.}(1999)Pistol, Castrillo, Hessman, Prieto, and
  Samuelson]{Pistol1999}
Pistol,~M.-E.; Castrillo,~P.; Hessman,~D.; Prieto,~J.~A.; Samuelson,~L.
  \emph{Phys. Rev. B} \textbf{1999}, \emph{59}, 10725--10729\relax
\mciteBstWouldAddEndPuncttrue
\mciteSetBstMidEndSepPunct{\mcitedefaultmidpunct}
{\mcitedefaultendpunct}{\mcitedefaultseppunct}\relax
\EndOfBibitem
\bibitem[Efros and Rosen(1997)Efros, and Rosen]{Efros1997}
Efros,~A.~L.; Rosen,~M. \emph{Phys. Rev. Lett.} \textbf{1997}, \emph{78},
  1110--1113\relax
\mciteBstWouldAddEndPuncttrue
\mciteSetBstMidEndSepPunct{\mcitedefaultmidpunct}
{\mcitedefaultendpunct}{\mcitedefaultseppunct}\relax
\EndOfBibitem
\bibitem[Shimizu \latin{et~al.}(2002)Shimizu, Woo, Fisher, Eisler, and
  Bawendi]{Shimizu2002}
Shimizu,~K.~T.; Woo,~W.~K.; Fisher,~B.~R.; Eisler,~H.~J.; Bawendi,~M.~G.
  \emph{Phys. Rev. Lett.} \textbf{2002}, \emph{89}, 117401\relax
\mciteBstWouldAddEndPuncttrue
\mciteSetBstMidEndSepPunct{\mcitedefaultmidpunct}
{\mcitedefaultendpunct}{\mcitedefaultseppunct}\relax
\EndOfBibitem
\bibitem[Park \latin{et~al.}(2007)Park, Link, Miller, Gesquiere, and
  Barbara]{Park2007}
Park,~S.-J.; Link,~S.; Miller,~W.~L.; Gesquiere,~A.; Barbara,~P.~F. \emph{Chem.
  Phys.} \textbf{2007}, \emph{341}, 169--174\relax
\mciteBstWouldAddEndPuncttrue
\mciteSetBstMidEndSepPunct{\mcitedefaultmidpunct}
{\mcitedefaultendpunct}{\mcitedefaultseppunct}\relax
\EndOfBibitem
\bibitem[Frantsuzov \latin{et~al.}(2008)Frantsuzov, Kuno, Jank\'{o}, and
  Marcus]{Frantsuzov2008}
Frantsuzov,~P.; Kuno,~M.; Jank\'{o},~B.; Marcus,~R.~A. \emph{Nat. Phys.}
  \textbf{2008}, \emph{4}, 519--522\relax
\mciteBstWouldAddEndPuncttrue
\mciteSetBstMidEndSepPunct{\mcitedefaultmidpunct}
{\mcitedefaultendpunct}{\mcitedefaultseppunct}\relax
\EndOfBibitem
\bibitem[Verberk \latin{et~al.}(2002)Verberk, van Oijen, and
  Orrit]{Verberk2002}
Verberk,~R.; van Oijen,~A.~M.; Orrit,~M. \emph{Phys. Rev. B} \textbf{2002},
  \emph{66}, 233202\relax
\mciteBstWouldAddEndPuncttrue
\mciteSetBstMidEndSepPunct{\mcitedefaultmidpunct}
{\mcitedefaultendpunct}{\mcitedefaultseppunct}\relax
\EndOfBibitem
\end{mcitethebibliography}

\end{document}